\pdfoutput=1
\documentclass[aps,twocolumn,groupedaddress]{revtex4-1}

\usepackage[T1]{fontenc}
\usepackage[latin9]{inputenc}
\usepackage{dsfont}
\usepackage{bbm}
\usepackage{braket}
\usepackage{amsmath}  
\usepackage{graphicx}
\usepackage{mathtools}

\begin{document}

\title{Observing the transition from quantum to classical energy correlations with photon pairs}
\author{Stefan Lerch, Andr\'e Stefanov}
\address{Institute of Applied Physics, University of Bern, 3012 Bern, Switzerland}

\email{andre.stefanov@iap.unibe.ch}

\begin{abstract}

We experimentally demonstrate a method to control the relative amount of quantum and classical energy correlations between two photons from a pair emitted by spontaneous parametric downconversion. Decoherence in the energy basis is achieved by applying random spectral phases on the photons. As a consequence a diverging temporal second order correlation function is observed and is explained by a mixture between an energy entangled pure state and a fully classically correlated mixte state. 

\end{abstract}
\maketitle

Quantum entanglement implies correlations beyond those allowed by classical models and plays a fundamental role in quantum metrology \cite{Giovannetti2011a}. In optical metrology, photon pairs generated by spontaneous parametric downconversion (SPDC) \cite{Rubin1994a} are the most practical realization of correlated light beams. Because the process is coherent, the emitted photons are in essence entangled, but it is of interest to experimentally be able to tune the correlations of quantum states, from purely quantum to classical. In particular since it is not always obvious to distinguish between advantages in metrology originating from genuine entanglement and effects due to classical correlations \cite{Stefanov2017}. Indeed, while some schemes where previously thought to be based on entanglement, they actually only rely on classical correlations. For instance in ghost imaging \cite{Pittman1995}, coincidence measurements from a thermal source allow to reproduce the image of an object, without the need of non-classical transverse correlations as originating from an SPDC source \cite{Valencia2005}. Another example can be found in photon number correlations. In SPDC, the down-converted photons are created pairwise, leading to a linear absorption rate for two-photon absorption processes \cite{Javanainen1990}. However, classical thermal light also shows photon bunching, and thus, can be exploited to enhance two-photon absorption as well \cite{Jechow2013}. The quantum nature of dispersion cancellation was also subject to debate \cite{Franson2009,Shapiro2010}.

Energy entangled biphoton states are an essential tool in the prospect of experimentally realizing quantum spectroscopy experiments \cite{Schlawin2013a,Dorfman2016}, and more generally for any energy-time two-photon metrology scheme, as for example quantum optical coherence tomography \cite{Abouraddy2002}.  Here the relevance of entanglement can also be misleading. For instance, in the case of two independent atoms, each of them excited by a single photon, a predicted enhanced absorption rate was first attributed to energy-time entanglement between the photons \cite{Muthukrishnan2004}. Yet it was later shown, that only classical frequency anticorrelations are actually enhancing the absorption rate \cite{Zheng2013}. Similarly in \cite{Schlawin2016} a pump-probe scheme is proposed, where a sample is excited by a classical pulse and probed by a photon from an entangled pair. Again, such scheme rely fundamentally on energy correlations and not on genuine entanglement.

In this paper we propose and experimentally realize a scheme to control transition from quantum to classical correlations with energy correlated photons. A characteristic of such entangled photon pairs is to show very strong energy correlations, together with strong temporal correlations. By introducing random phases on their spectral components and measuring the temporal shape of the two photon wavefunction, we observe a decrease of the temporal correlations. For polarization entangled photonic states, adding random phase has been shown to be a useful tool to generate in a controlled way mixed states required to test quantum protocols \cite{Kwiat2000,Altepeter2004,Bourennane2004,Prevedel2007,Xu2010b,Kim2011,Tsujimoto2015}. In the present work, the possibility to control the amount of classical versus quantum energy correlations opens the way for practical demonstration of the genuine advantage of entanglement, for instance as in quantum spectroscopy schemes \cite{Dorfman2016}.

Energy entangled photon pairs as emitted by SPDC are perfectly anti-correlated in energy when pumped by a monochromatic light. Their quantum state can be expressed as
\begin{equation}
\ket{\Psi}=\int\mathrm{d}\Omega\;\Lambda(\Omega)\ket{\Omega}_i\ket{-\Omega}_s,
\label{eq:quat_state}
\end{equation}
where $\ket{\Omega}_{i,s}$ is the state of an idler, respectively signal, photon with energy $\omega=\omega_p/2+\Omega$, and $\omega_p$ is the sum energy of both photons, being determined by the pump energy. This state also shows strong correlations in the time domain as can be observed from its expression in the time basis
\begin{equation}
\ket{\Psi}=\int\mathrm{d}\tau\;\hat{\Lambda}(\tau)\ket{\tau}_i\ket{\tau}_s,
\label{eq:quat_state_time}
\end{equation}
where $\hat{\Lambda}(\tau)$ is the Fourier transform of $\Lambda(\Omega)$ and $\ket{\tau}_{i,s}$ the state of a photon created at time $\tau$.

In order to demonstrate the real advantage of entanglement for quantum measurements, it is needed to be able to generate states of light where the energy correlations can be continuously tuned from purely quantum, as given by the state of Eq. (\ref{eq:quat_state}) to fully classical, all other parameters of the state being egal. The classically correlated state is described by a mixed state
\begin{eqnarray}
\hat{\rho}^{(c)}&=&\int\mathrm{d}\Omega\;p(\Omega)\ket{\Omega}_i\prescript{}{i}{\bra{\Omega}}\ket{-\Omega}_s\prescript{}{s}{\bra{-\Omega}},
\label{eq:T_rhoclassical}
\end{eqnarray}
with $p(\Omega)=\left|\Lambda(\Omega)\right|^2$.

The main difference between $\hat{\rho}^{(q)}=\ket{\Psi}\bra{\Psi}$ and $\hat{\rho}^{(c)}$ are the coherence terms between different frequencies. Assuming we can apply an arbitrary transfer function $M_j(\Omega)$ on the photons, the entangled state $\hat{\rho}^{(q)}$ transforms into $\hat{\rho}_j = \ket{\Psi_j}{\bra{\Psi_j}}$ with
\begin{equation}
\ket{\Psi_j}=\int\mathrm{d}\Omega\;\Lambda(\Omega)M_j(\Omega)M_j(-\Omega)\ket{\Omega}_i\ket{-\Omega}_s.
\label{eq:SwithM}
\end{equation}
In order to generate an arbitrarily correlated state, we can induce phase decoherence on  $\hat{\rho}^{(q)}$. This is realized by applying random transfer functions $M_j(\Omega)$ chosen from a set $\{M_j(\Omega)\}$, $j\in\{1,2,...,N\}$ with probabilities $p_j$. The state then becomes
\begin{eqnarray}
\hat{\rho}&=&\sum\limits_{j=1}^Np_j\hat{\rho}_j.
\label{eq:T_rhoens}
\end{eqnarray}
We select the transfer functions to be dephasing operations with random phases

\begin{equation}
M_j(\Omega)=\begin{cases} 
0, & \Omega< 0,\\
\mathrm{e}^{i\phi_j(\Omega)}, & \Omega\geq 0.
\end{cases}
\label{eq:T_randphi}
\end{equation}
The random variables $\phi_j(\Omega)$ follow a Gaussian distribution with average 0 and variance $\sigma^2$. They take not only a random value for different $j$ but also for different $\Omega$. The sum over all states in Eq. \eqref{eq:T_rhoens} can be evaluated as a sum over all transfer functions. For a sufficiently large mixture ($N\rightarrow\infty$), it is given by the correlation function
\begin{eqnarray}
\lefteqn{\braket{M_j(\Omega)M_j(-\Omega)M^*_j(\Omega')M^*_j(-\Omega')}=} \nonumber \\
&&\lim\limits_{N\rightarrow \infty}\sum\limits_{j=1}^Np_j\mathrm{exp}\left\{i\left[\phi_j(\Omega)+\phi_j(-\Omega)-\phi_j(\Omega')-\phi_j(-\Omega')\right]\right\}\nonumber\\
&&=\begin{cases}1, & |\Omega'|=|\Omega|\\ \mathrm{e}^{-\sigma^2}, & |\Omega'|\neq|\Omega|.\end{cases}
\label{eq:T_meanM}
\end{eqnarray}
Making use of Eq. \eqref{eq:T_meanM} in Eq. \eqref{eq:T_rhoens} leads to the final state
\begin{equation}
\hat{\rho}=\mathrm{e}^{-\sigma^2}\hat{\rho}^{(q)} +(1-\mathrm{e}^{-\sigma^2})\hat{\rho}^{(c)}.
\label{eq:T_transition}
\end{equation}
By tuning the variance $\sigma^2$ of the random phase distribution from zero to infinity, the final state undergoes a smooth transition from the entangled pure state $\hat{\rho}^{(q)}$ to a classically frequency anti-correlated state $\hat{\rho}^{(c)}$. 

\begin{figure}[h]
  \centering
\includegraphics[width=\columnwidth]{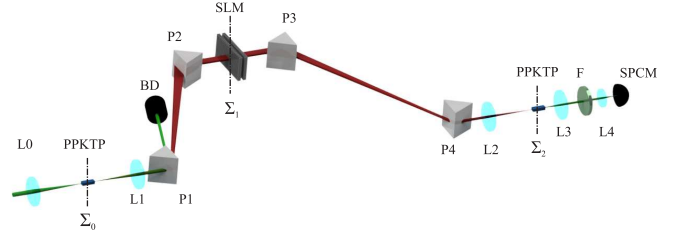} %\columnwidth
\caption{Schematic of the experimental setup. The pump laser is focused (L0) into a PPKTP crystal and generates energy-time entangled photons at plane $\Sigma_0$. The photons propagate through a four-prism compressor (P1-P4) and are imaged (L1,L2) to $\Sigma_2$. The spatially separated spectrum is shaped by an SLM at the symmetry plane $\Sigma_1$. The residual pump is blocked by a beam dump (BD). In a second PPKTP crystal, the photons are detected by means of up-conversion and the resulting upconverted photons are imaged by lenses L3 and L4 to a single photon counting module (SPCM), while the residual down-converted light is blocked by filters (F).}
\label{fig:D_setup}
\end{figure}

Figure 1 shows the experimental realization. A photon from a narrowband pump laser at 532 nm is downconverted into a pair of photons, idler and signal, each of them having a broad energy spectrum $\Lambda(\Omega)$ centred around 1064 nm with a width of about 40 nm, as measured by a spectrometer. The generated state is of the form of Eq. (\ref{eq:quat_state}). In order to verify that the photons are also temporally correlated according to Eq. (\ref{eq:quat_state_time}), the second order correlation function $G^{(2)}(\tau) $, proportional to  $\left|\hat{\Lambda}(\tau)\right|^2$, has to be measured by coincidence measurements with a time resolution shorter than its width. This is achieved by broadband up-conversion of both photons in a second non-linear crystal \cite{Dayan2005, ODonnell2009a}, and by applying the required transfer functions on the photons spectrum with the help of a pulse shaper \cite{Lerch2017}. The later is inspired from ultrafast optics and combines dispersive elements in a prism compressor configuration with a spatial light modulator (SLM). The dispersion introduced by the compressor is tuned by changing the position of the prisms. It is set such that the total dispersion induced by the optical setup from the source to the detection crystal is compensated. The fine tuning is performed by introducing a quadratic phase on the SLM. The width of the measured $G^{(2)}(\tau) $ is minimal when the dispersion is fully compensated. The entanglement in this configuration has been demonstrated by observing non-local dispersion compensation \cite{Lerch2017} and used for quantum information protocols \cite{Bernhard2013}. Explicitly, the second order correlation function is given by
\begin{equation}
G^{(2)}(\tau)=\left|\int\mathrm{d}\Omega\;\Lambda(\Omega)e^{i\Omega\tau}\right|^2.
\label{eq:G2simple}
\end{equation}
However, as only the even component of the transfer functions is relevant as seen from Eq. \eqref{eq:SwithM}, we have to express  $G^{(2)}(\tau)$ solely in terms of experimentally measurable quantities. Using the symmetry property $\Lambda(\Omega)=\Lambda(-\Omega)$, we derive the identity \cite{Lerch2017}
\begin{equation}
G^{(2)}(\tau)=\left|\int \mathrm{d}\Omega\;\Lambda(\Omega)e^{i|\Omega|\tau}\right|^2-4\left|\int\limits_0^\infty\mathrm{d}\Omega\;\Lambda(\Omega)\sin(\Omega\tau)\right|^2
\end{equation}
The first term in the right side can be implemented by a transfer function
\begin{equation}
M_a(\Omega,\tau)=\begin{cases} e^{-i \Omega \tau/2}, & \Omega<0\\ e^{i \Omega \tau/2}, & \Omega\geq 0,\end{cases}
\label{M1}
\end{equation}
and the second term by a transfer function 
\begin{equation}
M_b(\Omega,\tau)=\begin{cases}1, & \Omega<0\\ \sin(\Omega\tau), & \Omega \geq 0.\end{cases}
\label{M2}
\end{equation}
Therefore, $G^{(2)}(\tau)$ can be computed by taking the difference between two measured count rates $S(\tau) \propto S_{a}(\tau) - S_{b}(\tau)$. Here, $S_{a}(\tau) \doteq S[M_j(\Omega)M_{a}(\Omega,\tau)]$ is the signal measured with a total transfer function given by the product of the transfer function of Eq.~\eqref{M1} and the random function given by Eq.~\eqref{eq:T_randphi} inducing decoherence; similarly, $ S_{b}(\tau) \doteq 4 S[M_j(\Omega)M_{b}(\Omega,\tau)] $ with the transfer function of Eq.~\eqref{M2}.

Each of the signals needed to measure the time difference between signal and idler are averaged over 100 acquisitions of one second each with different random transfer functions implemented on the SLM, such that for $\quad k\in\{a,b\}$
\begin{equation}
\resizebox{\linewidth}{!}{$\braket{S_k(\tau,\sigma)}\propto\left\langle\left|\int\mathrm{d}\Omega\;\Lambda(\Omega)M_k(-\Omega,\tau)M_k(\Omega,\tau)\mathrm{e}^{i(\phi_j(\Omega)+\phi_j(-\Omega)}\right|^2\right\rangle$} 
\end{equation}
Because of the linearity of the trace, it is valid to evaluate the signal contribution of each transfer function setting separately and average over all $j$ afterwards. The procedure is repeated for different values of $\sigma$.

\begin{figure}[hb]
\includegraphics[width=\columnwidth]{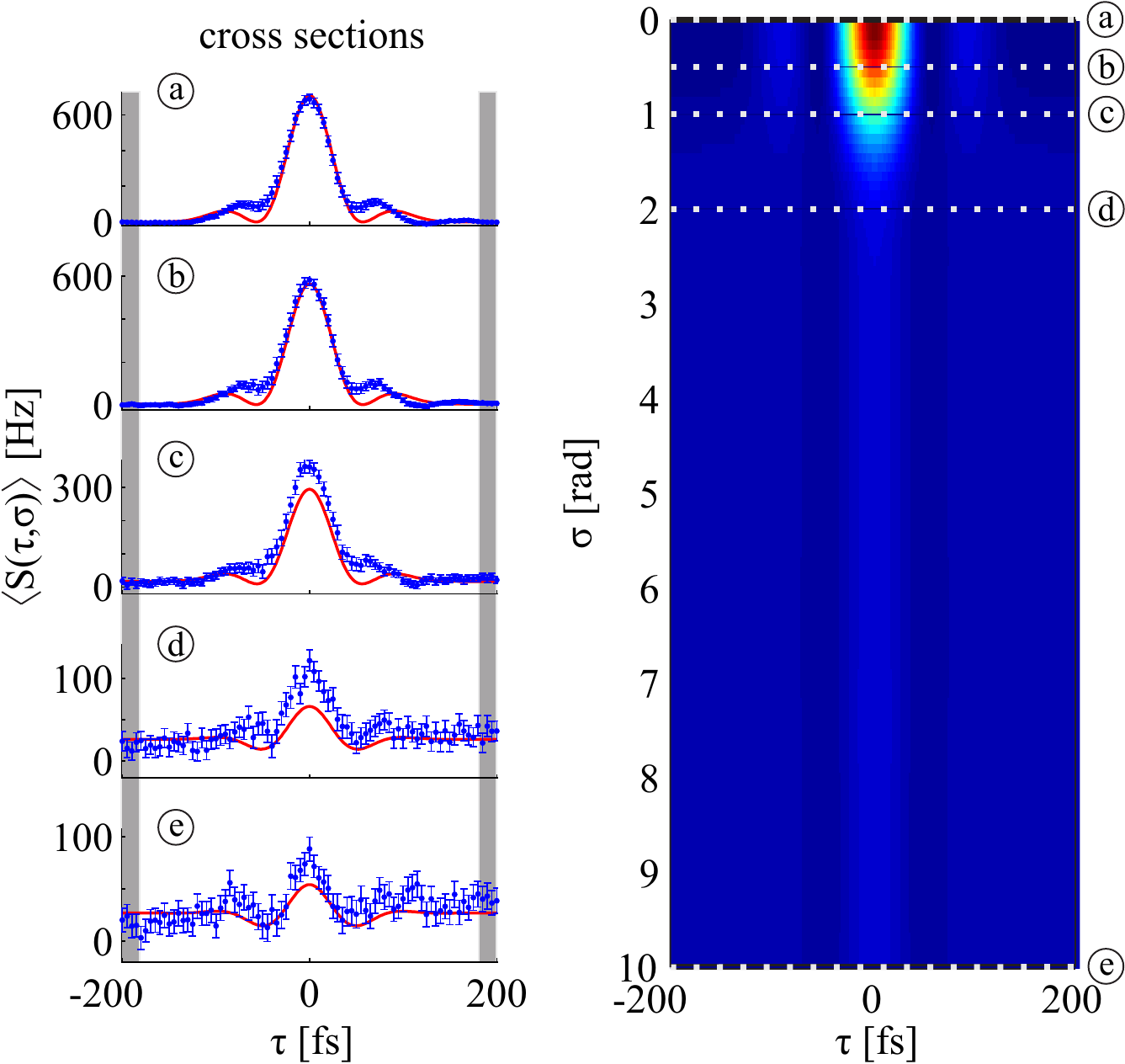}
\caption{left: Time of arrival difference between signal and idler $S(\tau)$ for increasing amount of phase noise $\sigma$ from top to bottom: $\sigma=0,05,1,2,10$ rad. Dark count subtracted measurements (blue dots) and model (red curve). The shaded gray region labels the averaging region that is used to estimate the background level.
\\
right: Simulated time of arrival difference between signal and idler $\braket{S(\tau,\sigma)}$ as a function of noise. The white dotted lines, labelled with letters a to d, indicate the cross sections, where measurements are taken.
}
\label{fig:T_results}
\end{figure}
The transfer function of the SLM is linear and disentangles the state by introducing phase decoherence \cite{Zurek2003}. As a consequence, we are able to reduce the temporal correlations without changing the energy correlations. 

Figure \ref{fig:T_results} shows the results of the measured two photon temporal correlation functions for various noises $\sigma$, together with curves computed from a model of the density matrix of the state. The error bars are calculated under the assumptions of a Poisson distribution. The parameters of the theory are determined by fitting the measurement with $\sigma=0$ rad according to a model of the signal given by ${S(\tau,\sigma=0)}= B \cos\left(\mu\tau\right)\mathrm{e}^{-\sigma'^2\tau^2/2}$, with the fitting parameters $B$, $\mu$ and $\sigma'$. This model is justified by the fact that the entangled photon spectrum for the chosen SPDC phase matching can be well approximated by a double Gaussian curve given by
\begin{equation}
S(\omega,\sigma=0)= \frac{B}{2\sigma'}\left(e^{- \frac{(\omega-\mu)^2}{2\sigma'^2}}+e^{- \frac{(\omega+\mu)^2}{2\sigma'^2}}\right).
\end{equation}
We find $B=708.71$ Hz, $\sigma'=0.022$ rad/fs, and $\mu = 0.0275$ rad/fs. The corresponding spectral width given by $\sqrt{\sigma'^2+\mu^2}$ is 21 nm. It is smaller than the width measured with a spectrometer, as only a part of the SPDC spectrum is contributing to the up-conversion signal. In the present experimental configuration it is not possible to measure directly the width of the effectively up-converted spectrum and therefore it is a fitting parameter of the temporal measurement.

\begin{figure}[hb]
\includegraphics[width=\columnwidth]{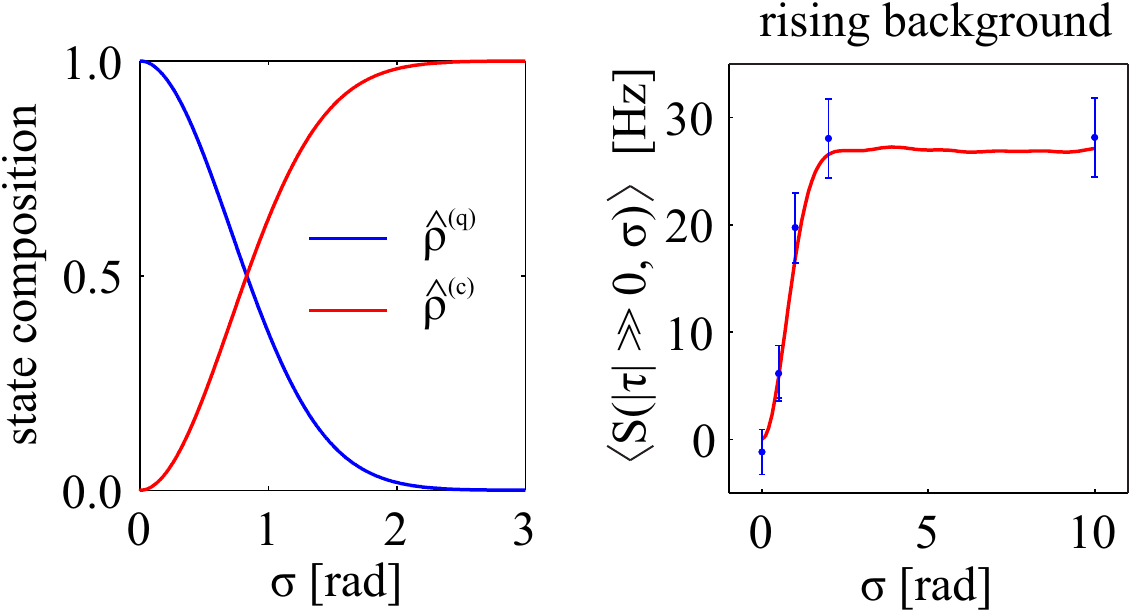}
\caption{left: Calculated fraction of classically correlated $\hat{\rho}^{(c)}$ and purely entangled states $\hat{\rho}^{(q)}$ in the mixture as a function of $\sigma$.
\\
right: Background of $\braket{S(\tau,\sigma)}$ as a function of $\sigma$}
\label{fig:T_results2}
\end{figure}

We then evaluate Eq.~\eqref{eq:SwithM} for 10000 random SLM settings and simulate the expected signals. They are then averaged in order to compute the correlation function for arbitrary noise $\sigma$, as shown on Fig. \ref{fig:T_results}. It should be noted that we don't observe a broadening of the correlation peak, as it would be expected simply from dispersion, but a mixture of two features in the temporal curves. Apart from a decreasing correlation peak whose shape remains constant, we observe an increasing constant background. This is the component leading to a diverging time-difference variance, as pointed out in \cite{Wasak2010}. The model allows to compute the fractions of entangled state and classically correlated state as seen on Fig. \ref{fig:T_results2} left. They are equal for $\sigma=\sqrt{\ln(2)}\approx 0.833$ rad. In order to further estimate the relative contributions of those two components of the signal, we evaluate the mean of the signal $\braket{S(\tau,\sigma)}$ in the range of $\tau\in\{[-200,-180],[180,200]\}$ fs, indicated in Fig. \ref{fig:T_results} by the gray shaded region. For $\sigma=2$ rad the mixture consists of less than 2\% entangled states, and thus the background reaches its asymptotic value, given by $(26.93\pm 0.03)$ Hz for the simulation and $(28.8\pm 1.8)$ Hz for the measurement. The simulation follows nicely the measured values (Fig. \ref{fig:T_results2} right). The rising background is proportional to the fraction of $\hat{\rho}^{(c)}$ in Eq. \eqref{eq:T_transition} that contributes to the measured mixture.

In conclusion, we have demonstrated the full control on the degree of quantum  versus classical energy correlations in photon pairs by adding random phase noise on the photons spectrum. We have observed a reduction of the strength of the temporal correlations and their divergence in agreement with the theory. Such a tunable entangled source is an essential tool to experimentally verify the fundamental advantage of entangled states against classical correlations in any setup relying on time-energy measurements.

\begin{acknowledgements}
This research was supported by the Swiss National Science Foundation through the grant PP00P2\_159259.
\end{acknowledgements}

\bibliographystyle{apsrev4-1} 
%\bibliography{C:/Users/Andre/Documents/library}
%
\end{document}